\documentclass[pra,aps,10pt,twocolumn,showpacs,amsmath,amssymb]{revtex4-1}

\usepackage[english]{babel}
\usepackage{afterpage}
\usepackage{graphicx}
\usepackage{dcolumn}
\usepackage{bm}
\usepackage{color}
\usepackage{slashed}
\usepackage{latexsym}

\newcommand{\dd}{d}  
\newcommand{\ii}{i}  
\newcommand{\ee}{e}  

\begin{document}

\title{Generation of attosecond electron beams in relativistic ionization by short laser pulses}
\author{F. Cajiao V\'elez}
\author{J. Z. Kami\'nski}
\author{K. Krajewska}
\email[E-mail address:\;]{Katarzyna.Krajewska@fuw.edu.pl}
\affiliation{Institute of Theoretical Physics, Faculty of Physics, University of Warsaw, Pasteura 5,
02-093 Warsaw, Poland}
\date{\today}

\begin{abstract}
Ionization by relativistically intense short laser pulses is studied in the framework of strong-field quantum electrodynamics.
Distinctive patterns are found in the energy probability distributions of photoelectrons. Except of the already observed patterns, which were studied in Phys. Rev. A {\bf 94}, 013402 (2016),
we discover an additional interference-free smooth supercontinuum in the high-energy portion of the spectrum, reaching tens of kiloelectronovolts. As we show, the latter is sensitive to the driving field intensity and it
can be detected in a narrow polar-angular window. Once these high-energy electrons are collected, they can form solitary attosecond pulses. This is particularly important in light of various applications of
attosecond electron beams such as in ultrafast electron diffraction and crystallography, or in time-resolved electron microscopy of physical, chemical, and biological processes.
\end{abstract}


\maketitle

\section{Introduction}
\label{sec::intro}

In recent years, the synthesis of electron wave packets of short duration has attracted a lot of attention due to its applications in ultrafast electron diffraction, 
crystallography, and microscopy~\cite{Srinivasan2003,Zewail2005,lobastov,Zewail2006}. It has been shown that femtosecond electron pulses can be used to image complex 
molecular, biological, and crystalline structures with very short temporal resolution. Therefore, it has become possible
to observe transient molecular structures in the course of chemical reactions~\cite{Ihee,Lobastov2001,Srinivasan,Gao,Centurion},
gather an insight into melting and heating processes~\cite{Siwick,Ernstorfer,Sciaini}, or to observe phase transitions in crystalline and polycrystalline 
materials~\cite{Gedik2007,baum}.

Typically, short electron wave packets are produced by shining laser pulses of short duration upon flat photocathodes~\cite{Aidelsburger2010} or sharp metallic 
tips~\cite{Hommelhoff2006b,Ropers2007,Schenk2010,Hommelhoff2006,Kruger2011}. Other techniques include electron emission and acceleration in intense plasmon 
fields~\cite{Kupersztych2001,Irvine2004,Irvine2006,Greig2014,Greig2016}, and photoemission from supercooled atoms in optical traps~\cite{Claessens2005,Taban2010,Engelen2013}.
Photoelectrons obtained by these methods appear in vacuum with certain velocity distributions and need to be accelerated up to energies appropriate for microscopy 
or diffraction. During acceleration and posterior free-propagation, the wave packets spread in time. Additionally, if the short pulses contain several electrons, 
Coulomb repulsion among them contributes importantly to the temporal broadening~\cite{Siwick2002,Paarmann2012}. 
Thus, in order to avoid space-charge effects, synthesis of ultrashort pulses consisting of single electrons has been considered~\cite{Aidelsburger2010,Sciaini2011,zewail,lobastov,baum2,baum3,Baum2013}.

The shortest single electron pulses with a full width at half maximum duration of 28~fs have been demonstrated recently~\cite{Gliserin}. Moreover, various proposals for further compression
of these pulses to attosecond durations have been put forward~\cite{baum2,baum3,Fill,Hilbert,Gliserin2012,Hansen}. 
At this point, it is also important to note that theoretical investigations of scattering processes employing attosecond electron pulses have confirmed their ability 
to image electronic motions in target atoms and molecules~\cite{Shao1,Shao2,Shao3,Shao4}.

Yet, another proposal for producing attosecond electron pulses
has been introduced in~\cite{no_interference}. In this paper, we have shown that very short electron bunches, with energies of few keV, can be produced by
the interaction of hydrogen-like ions with intense circularly-polarized laser fields. In this case, a broad structure located at the high-energy portion of the photoelectron
spectrum is formed (see, also Ref.~\cite{proceedings_no_interference}). This structure, the so-called supercontinuum, is characterized by the absence of multiphoton interferences. 
Moreover, it does not present significant fluctuations of probability in the range of tens to hundreds single-photon energies and, according to
the space-time analysis, can be used to obtain electron pulses with attosecond duration.

It is the aim of this paper to further analyze the properties of the supercontinuum in photoionization and its
application to the generation of attosecond electron wave packets. First, we show that the photoelectron energy spectra can exhibit not just one, 
but two well-defined interference-free structures at different photoelectron kinetic energies. For the laser field parameters
considered here and for the fixed electron detection angles, the high-energy structure is approximately two orders of magnitude more pronounced
than the mid-energy one and, hence, it contributes the most to the energy-integrated probability distribution at fixed angles. Next, we show that
this structure can be used to synthesize attosecond electron pulses. As we demonstrate, the time duration of 
the resulting electron pulses can be shortened by increasing the intensity of the laser field which drives ionization.

In contrast to our previous works~\cite{no_interference,proceedings_no_interference}, we consider here ionization stimulated by low-frequency laser fields. 
This has been motivated by the fact that such fields, with high intensities and short durations, 
will soon be available in laser facilities such as Extreme Light Infrustructure (ELI)~\cite{ELI} or Exawatt Center for Extreme Light Studies (XCELS)~\cite{XCELS}.
Moreover, it appears beneficial to use
low-frequency incident laser fields in the context of electron pulse generation. The point is that, if we fix the laser field intensity, the lower its frequency the more energetic 
photoelectrons will be detected (since the ponderomotive energy of electron oscillations in a laser field increases). This, in turn, will lead to the synthesis of shorter electron pulses.

We would like to point out that the synthesis of attosecond electron pulses from the above-mentioned high-energy supercontinuum
may have at least two advantages. First, the detected photoelectrons have velocities appropriate for ultrafast
electron microscopy and diffraction, which makes their further acceleration unnecessary.
Second, those photoelectrons are observed within a narrow angular window and, depending on the actual size of the final electron beam, additional collimation
methods might be less needed. Note, however, that ultrashort electron wave packets have large energy bandwidths, as a
direct consequence of the uncertainty principle. It is, therefore, expected that nonrelativistic short electron pulses
spread fast in time during free propagation. This is not the case for wave packets synthesized with relativistic kinetic
energies (close to MeV), as their natural spreading is generally slower. We show here that very energetic photoelectrons
can actually be obtained by employing laser fields of larger intensity. Nevertheless, for electron microscopy
applications, standard compression techniques can also be applied~\cite{Varin,baum2,baum3,Oudheusden2007,Veisz2007,Oudheusden2010,Chatelain2012}.

This paper is organized as follows. 
In Sec.~\ref{ionization_probability_distribution}, for convenience of the reader, we present the formulas of the photoionization probability distributions under the scope of the relativistic 
strong-field approximation and the plane-wave front approximation. The respective derivations were originally introduced in Ref.~\cite{no_interference}. While Sec.~\ref{energy_prob} 
relates to the energy spectra of photoelectrons and the formation of the supercontinuum, Sec.~\ref{wave_packets} is dedicated to the analysis of the space-time probability distributions 
and the formation of attosecond electron wave packets. Also in Sec.~\ref{wave_packets}, the validity of the plane-wave front approximation for the driving laser pulse is tested for our calculations. Finally, 
in Sec.~\ref{conclusions} we summarize our results and outline the perspectives for further investigations.

\section{Theoretical formulation}
\label{ionization_probability_distribution}

Consider a hydrogen-like ion interacting with a relativistically strong laser pulse. The probability amplitude of ionization from the initial bound state of energy $E_0$, $\Psi_{\mathrm{i}}(x)$, 
to the final scattering state, ${\Psi}_{\mathrm{f}}(x)$, is given as~\cite{reiss_RSFA,no_interference, proceedings_no_interference}
\begin{equation}
\mathcal{A}_{\mathrm{fi}}=-\ii \int\dd^4x \ee^{-\ii (E_0/c)x^0}\bar{\Psi}_{\mathrm{f}}(x)e\slashed{A}_{\mathrm{R}}(x)\Psi_{\mathrm{i}}(\bm{x}),
\label{theory1}
\end{equation}
where the four-vector ${A}_{\mathrm{R}}(x)$ represents the electromagnetic potential describing the laser field and $e<0$ is the electron charge. 
While Eq.~\eqref{theory1} is exact and $\Psi_{\mathrm{i}}(x)$ can be derived analytically for the Coulomb potential (see, e.g., Ref.~\cite{BjorkenDrell}), 
${\Psi}_{\mathrm{f}}(x)$ can only be determined numerically for laser fields of moderate intensities. For this reason, in the relativistic 
strong-field approximation (RSFA), the exact scattering state $\Psi_{\rm f}(x)$ is usually replaced by the Volkov solution~\cite{Volkov} (i.e, the solution of the Dirac equation 
coupled to the laser field). Therefore, in the RSFA, the spin-fixed probability amplitude of ionization, now denoted as $\mathcal{A}_{\lambda\lambda_{\rm i}}(\bm{p})$, takes the form,
\begin{align}
\mathcal{A}_{\lambda\lambda_{\rm i}}(\bm{p})=&-\ii \int\frac{\dd^3q}{(2\pi)^3}\int\dd^4x\, \ee^{-\ii q\cdot x}\bar{\psi}^{(+)}_{\bm{p}\lambda}(x)\nonumber\\
\times&e\slashed{A}_{\mathrm{R}}(x)\tilde{\Psi}_{\mathrm{i}}(\bm{q}).
\label{theory2}
\end{align}
Here, ${\psi}^{(+)}_{\bm{p}\lambda}(x)$ is the Volkov solution describing the electron with an asymptotic momentum ${\bm p}$ and spin polarization $\lambda=\pm$. 
$\tilde{\Psi}_{\mathrm{i}}(\bm{q})$ is the Fourier transform of the bound state ${\Psi}_{\mathrm{i}}(\bm{x})$ and $\lambda_{\mathrm{i}}$ is the initial electron spin. 
Note that, in Eq.~\eqref{theory2}, we have introduced $q=(q^0,\bm{q})=(E_0/c,\bm{q})$, which is not a four-vector as it does not transform properly under Lorentz transformations. 
Nevertheless, it will help us to simplify our further notation. Here, it is also worth noting that the RSFA is restricted to the case when the kinetic energy of photoelectrons is much larger than the ionization potential of the initial bound state, i.e., 
$\sqrt{(m_{\mathrm{e}}c^2)^2+(c\bm{p})^2}-m_{\mathrm{e}}c^2\gg m_\mathrm{e}c^2-E_0$, as discussed in Refs.~\cite{no_interference,proceedings_no_interference}.

Our calculations are carried out in the velocity gauge and the laser field is described using the plane-wave front approximation. In this case, 
the electromagnetic potential $A_{\rm R}(x)$ can be written as
\begin{equation}
A_{\mathrm{R}}(x)=A_0[\varepsilon_1 f_1(k\cdot x)+\varepsilon_2 f_2(k\cdot x)],
\label{theory3}
\end{equation}
where $k=k^0n=k^0(1,\bm{n})$ is the wave four-vector and $k^0=\omega/c$. In our notation, the unitary vector $\bm{n}$ represents the direction 
of propagation of the laser pulse, $\omega=2\pi/T_{\mathrm{p}}$ is its fundamental frequency, whereas $T_{\mathrm{p}}$ represents its duration. Additionally, 
$\varepsilon_j\equiv (0,{\bm{\varepsilon}}_j)$ ($j=1,2$) are two real and normalized polarization four-vectors perpendicular to the laser field propagation direction  
(i.e., $k\cdot\varepsilon_j=-{\bm k}\cdot{\bm \varepsilon}_j=0$). In Eq.~\eqref{theory3}, $f_j(\phi)$ represent two real shape functions with continuous second derivatives 
which vanish for $\phi<0$ and $\phi>2\pi$.

We recall that the Volkov solution for an electron, normalized in the volume $V$, equals~\cite{Volkov,FKK,PiazzaRev}
\begin{align}
\psi^{(+)}_{\bm{p}\lambda}(x)=&\sqrt{\frac{m_{\mathrm{e}}c^2}{VE_{\bm{p}}}}\Bigl(1+\frac{m_{\mathrm{e}}c\mu}{2p\cdot k}\bigl[f_1(k\cdot x)\slashed{\varepsilon}_1\slashed{k} \nonumber \\
& + f_2(k\cdot x)\slashed{\varepsilon}_2\slashed{k} \bigr] \Bigr) \ee^{-\ii S_p^{(+)}(x)}u^{(+)}_{\bm{p}\lambda},
\label{theory4}
\end{align}
where $m_{\rm e}$ is the electron mass, $p=(E_{\bm p}/c,{\bm p})$ is its asymptotic on-shell four-momentum, whereas
\begin{align}
S_p^{(+)}(x)&= p\cdot x+\int_0^{k\cdot x}\dd\phi\Bigl[-\frac{m_{\mathrm{e}}c\mu}{p\cdot k}\bigl(\varepsilon_1\cdot p f_1(\phi) \nonumber \\
& + \varepsilon_2\cdot p f_2(\phi)\bigr)+\frac{(m_{\mathrm{e}}c\mu)^2}{2p\cdot k}\bigl(f_1^2(\phi)+f_2^2(\phi)\bigr)\Bigr].
\label{theory5}
\end{align}
Moreover, $\mu=|eA_0|/(m_{\rm e}c)$ denotes the normalized amplitude of the vector potential~\eqref{theory3} 
whereas $u^{(+)}_{{\bm p}\lambda}$ is the free electron bispinor~\cite{BjorkenDrell}. Using these expressions,
we derive that the probability amplitude of ionization under the RSFA~\eqref{theory2} becomes
\begin{equation}
\mathcal{A}_{\lambda\lambda_{\rm i}}(\bm{p})=\int \frac{\dd^3q}{(2\pi)^3}\int\dd^4x \ee^{\ii S_p^{(+)}(x)-\ii q\cdot x} M_{\lambda\lambda_{\rm i}}(k\cdot x),
\label{theory6}
\end{equation}
where
\begin{align}
M_{\lambda\lambda_{\rm i}}&(k\cdot x)=\ii m_{\mathrm{e}}c\mu\sqrt{\frac{m_{\mathrm{e}}c^2}{VE_{\bm{p}}}}\nonumber\\
&\times \bigl[f_1(k\cdot x)B^{(1,0)}_{\bm{p}\lambda;\lambda_{\mathrm{i}}}(\bm{q})+f_2(k\cdot x)B^{(0,1)}_{\bm{p}\lambda;\lambda_{\mathrm{i}}}(\bm{q})\nonumber\\
&-\frac{m_{\mathrm{e}}c\mu}{2p\cdot n}\bigl([f_1(k\cdot x)]^2+[f_2(k\cdot x)]^2\bigr)B^{(0,0)}_{\bm{p}\lambda;\lambda_{\mathrm{i}}}(\bm{q})\bigr].
\label{theory7}
\end{align}
To simplify our notation, we have introduced the following functions expressed in terms of the Fourier transform of the initial ground state, $\tilde{\Psi}_{\rm i}({\bm q})$,
\begin{align}
B^{(0,0)}_{\bm{p}\lambda;\lambda_{\mathrm{i}}}(\bm{q})=&\bar{u}^{(+)}_{\bm{p}\lambda}\slashed{n}\tilde{\Psi}_{\mathrm{i}}(\bm{q}),\nonumber\\
B^{(1,0)}_{\bm{p}\lambda;\lambda_{\mathrm{i}}}(\bm{q})=&\bar{u}^{(+)}_{\bm{p}\lambda}\slashed{\varepsilon}_1\tilde{\Psi}_{\mathrm{i}}(\bm{q}),\nonumber\\
B^{(0,1)}_{\bm{p}\lambda;\lambda_{\mathrm{i}}}(\bm{q})=&\bar{u}^{(+)}_{\bm{p}\lambda}\slashed{\varepsilon}_2\tilde{\Psi}_{\mathrm{i}}(\bm{q}). \label{theory8}
\end{align}

In the next step, we define the so-called laser-dressed four-momentum of the electron, $\bar{p}$,
\begin{align}
\bar{p}=&p-\frac{m_{\mathrm{e}}c\mu}{p\cdot k}(\varepsilon_1\cdot p\langle f_1\rangle+\varepsilon_2\cdot p\langle f_2\rangle )k\nonumber\\
+&\frac{(m_{\mathrm{e}}c\mu)^2}{2p\cdot k}(\langle f_1^2\rangle+\langle f_2^2\rangle )k,
\label{theory9}
\end{align}
where the time averaging $\langle ...\rangle$ is over the pulse duration, $T_p$. As a result, the function $G_p(k\cdot x)$ that is periodic in $k\cdot x$ [meaning that 
$G_p(0)=G_p(2\pi)=0$] can be extracted in Eq.~\eqref{theory5}, $S_p^{(+)}(x)=\bar{p}\cdot x+G_p(k\cdot x)$. For completeness, we write it down,
\begin{align}
G_p(\phi)&=\int_0^\phi\dd\phi' \Bigl[-\frac{m_{\mathrm{e}}c\mu}{p\cdot k}\bigl(\varepsilon_1\cdot p (f_1(\phi')-\langle f_1\rangle) \nonumber \\
& + \varepsilon_2\cdot p (f_2(\phi')-\langle f_2\rangle)\bigr)+\frac{(m_{\mathrm{e}}c\mu)^2}{2p\cdot k}\bigl(f_1^2(\phi') \nonumber\\
& - \langle f_1^2\rangle + f_2^2(\phi')-\langle f_2^2\rangle\bigr)\Bigr].
\label{theory10}
\end{align}
Now, it is useful to introduce the light-cone coordinates in Eq.~\eqref{theory6}; namely, $x^-=x^0-{\bm n}\cdot {\bm x}, x^+=\frac{1}{2}(x^0+{\bm n}\cdot {\bm x})$, and 
${\bm x}^\perp={\bm x}-({\bm n}\cdot {\bm x}){\bm n}$. Since the pulse phase $k\cdot x=k^0x^-$, most of the integrals in Eq.~\eqref{theory6} become trivial. The only nontrivial 
integral over $x^-$ is performed with the help of the Fourier expansions~\cite{KKCompton,KKBW},
\begin{align}
\bigl[f_1(\phi)\bigr]^j\exp[\ii G_p(\phi)]=&\sum_{N=-\infty}^{\infty}G_N^{(j,0)}\ee^{-\ii N\phi},\nonumber \\
\bigl[f_2(\phi)\bigr]^j\exp[\ii G_p(\phi)]=&\sum_{N=-\infty}^{\infty}G_N^{(0,j)}\ee^{-\ii N\phi},\label{theory11}
\end{align} 
where $j=1,2$. Hence, the spin-resolved probability amplitude of ionization can be represented as an infinite sum,
\begin{equation}
\mathcal{A}_{\lambda\lambda_{\rm i}}(\bm{p})=\ii m_{\mathrm{e}}c\mu\sqrt{\frac{m_{\mathrm{e}}c^2}{VE_{\bm{p}}}}\mathcal{D}(\bm{p},\lambda;\lambda_{\mathrm{i}}),
\label{theory12}
\end{equation}
where
\begin{align}
\mathcal{D}(\bm{p},\lambda;\lambda_{\mathrm{i}})&=\sum_{N=-\infty}^{\infty}\frac{\ee^{2\pi\ii(\bar{p}^+-q^+-Nk^0)/k^0}-1}{\ii (\bar{p}^+-q^+-Nk^0)} \nonumber\\
&\times\Bigl\{G^{(1,0)}_N B^{(1,0)}_{\bm{p}\lambda;\lambda_{\mathrm{i}}}(\bm{Q})+ G^{(0,1)}_N B^{(0,1)}_{\bm{p}\lambda;\lambda_{\mathrm{i}}}(\bm{Q}) \nonumber \\
&-\frac{m_{\mathrm{e}}c\mu}{2p\cdot n}[G^{(2,0)}_N+G^{(0,2)}_N]B^{(0,0)}_{\bm{p}\lambda;\lambda_{\mathrm{i}}}(\bm{Q})\Bigr\}\label{theory13}
\end{align}
and $\bm{Q}=\bm{p}+(q^0-p^0)\bm{n}$.

Finally, integrating $|\mathcal{A}_{\lambda\lambda_{\rm i}}(\bm{p})|^2$
over the density of final electron states, $V\dd^3p/(2\pi)^3$, we obtain that the total probability of ionization equals
\begin{equation}
P_{\rm ion}=\mu^2\frac{(m_{\mathrm{e}}c)^3}{2(2\pi)^3}\sum_{\lambda,\lambda_{\rm i}=\pm}\int\frac{\dd^3p}{p^0}|\mathcal{D}(\bm{p},\lambda;\lambda_{\mathrm{i}})|^2.
\label{theory15}
\end{equation}
Here, the averaging over the initial and summation over the final electron spin degrees of freedom have been performed. Furthermore, the above equation
allows us to define the spin-resolved triply-differential probability distribution of ionization,
\begin{equation}
\frac{\dd^3P(\bm{p},\lambda;\lambda_{\mathrm{i}})}{\dd E_{\bm{p}}\dd^2\Omega_{\bm{p}}}=\mu^2\frac{(m_{\mathrm{e}}c)^3}{(2\pi)^3c}|\bm{p}|\cdot|\mathcal{D}(\bm{p},\lambda;\lambda_{\mathrm{i}})|^2.
\label{theory16}
\end{equation}
For the purpose of our numerical illustrations, we introduce also the dimensionless distributions,
\begin{equation}
\mathcal{P}_{\lambda\lambda_{\rm i}}(\bm{p})=\alpha^2m_{\mathrm{e}}c^2\frac{\dd^3P(\bm{p},\lambda;\lambda_{\mathrm{i}})}{\dd E_{\bm{p}}\dd^2\Omega_{\bm{p}}}
\label{theory17}
\end{equation}
and
\begin{equation}
{\mathcal P}({\bm p})=\frac{\alpha^2m_{\mathrm{e}}c^2}{2}\sum_{\lambda,\lambda_{\rm i}=\pm}\frac{\dd^3P(\bm{p},\lambda;\lambda_{\mathrm{i}})}{\dd E_{\bm{p}}\dd^2\Omega_{\bm{p}}},
\label{theory18}
\end{equation}
where $\alpha=e^2/(4\pi\varepsilon_0c)$ is the fine-structure constant. These are the probability distributions of ionization expressed in atomic units~\cite{no_interference}.

\subsection{Laser Pulse}
\label{sec:laser_pulse}

To illustrate the theory presented above, we consider the interaction of an intense and circularly polarized laser pulse with He$^+$ ions.
The latter are one-electron ions, with the atomic number $Z=2$ and the ionization potential of roughly 54~eV.
The laser pulse~\eqref{theory3} is characterized by the shape functions $f_j(\phi)$, for $j=1,2$, defined as
\begin{equation}
f_j(\phi)=-\int_0^{\phi}\dd \phi' F_j(\phi'),
\label{theory19}
\end{equation}
where
\begin{equation}
F_j(\phi)=N_0\sin^2\Bigl(\frac{\phi}{2}\Bigr)\sin(N_{\mathrm{osc}}\phi+\delta_j)\cos(\delta+\delta_j)
\label{theory20}
\end{equation}
for $0<\phi <2\pi$ and it is 0 otherwise. Here, $N_{\rm osc}$ is the number of cycles comprising the pulse, $N_0=\sqrt{\frac{8}{3}}N_{\rm osc}$ is a normalization constant which guarantees that the average 
intensity of the field is independent of the number of cycles~\cite{no_interference}, whereas $\delta$ and $\delta_j$ determine the polarization properties of the pulse.
Unless otherwise stated, in the following we consider a three-cycle ($N_{\rm osc}=3$) circularly polarized laser pulse ($\delta_1=0$, $\delta_2=\pi/2$, and $\delta=\pi/4$) that 
propagates along the $z$-axis ($\bm{n}={\bm e}_z$), with the polarization vectors ${\bm \varepsilon}_1={\bm e}_x$ and ${\bm \varepsilon}_2={\bm e}_y$. 
The laser carrier frequency, $\omega_{\mathrm{L}}=N_{\mathrm{osc}}\omega$, is equal to $1.5498$~eV.

\begin{figure}
\begin{center}
\includegraphics[width=5.2cm]{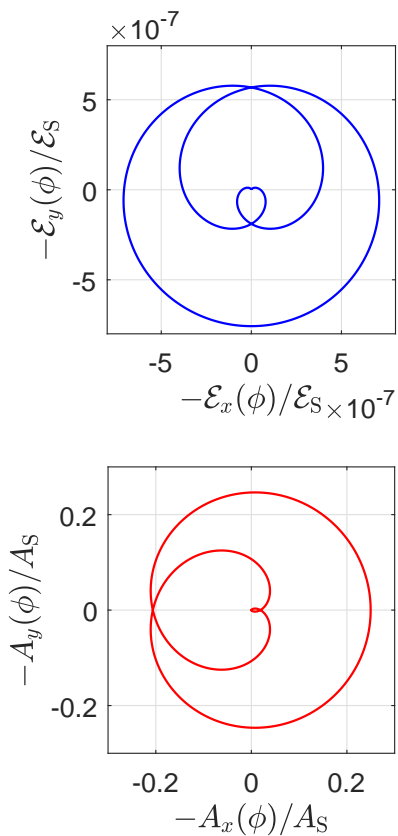}
\end{center}
\caption{Trajectories of the tips of the electric field vector $\bm{\mathcal{E}}(\phi)$ (upper panel) and the electromagnetic vector potential $\bm{A}(\phi)$ (lower panel) 
in relativistic units for the laser pulse parameters discussed in the paper. The time-averaged intensity 
is $I=10^{17}$~W/cm$^2$. All trajectories start and end up at the origin $(0,0)$, evolving counterclockwise. 
\label{figLaserFunctions160812}}
\end{figure}

In Fig.~\ref{figLaserFunctions160812}, we present the trajectories of the tips of the electric field vector $\bm{\mathcal{E}}(\phi)$ (upper panel) and the  vector potential 
$\bm{A}(\phi)$ (lower panel) in the $xy$-plane for the laser pulse described above. Both curves start and end up at the origin of coordinates and evolve counterclockwise. 
The parameter $\mathcal{E}_S=m_{\mathrm{e}}^2c^3/|e|$ is the so-called Sauter-Schwinger critical field~\cite{Sauter,Schwinger}, whereas $A_S=m_{\mathrm{e}}c/|e|$. 
Note that, for the laser field parameters considered in this paper, we have $|\bm{\mathcal{E}}(\phi)|/\mathcal{E}_{\mathrm{S}}\ll 1$. This implicates that, under current conditions,
the electron-positron pair creation from vacuum is negligible~\cite{no_interference}.

\subsection{Monte Carlo analysis}

In Sec.~\ref{ionization_probability_distribution} we have defined the triply-differential probability distribution of ionization of hydrogen-like ions 
interacting with short laser pulses [Eq.~\eqref{theory16}]. In this case, the initial ground-state wave function is unperturbed at times prior to the interaction with the laser field, 
independently of its intensity. This allows us to analyze the photoionization of light ions (such as He$^+$) by intense laser pulses. In contrast, 
when the infinite plane-wave approximation is considered (see, e.g., Refs.~\cite{eikHeidelberg,Klaiber,eikHeidelberg2,reiss_RSFA}), the RSFA is restricted to the case 
of highly charged positive ions interacting with laser fields of moderate intensity. This guarantees that the initial bound state is not heavily distorted by 
the action of the laser field.

Note also that, when a slowly-varying envelope is used to model the driving laser pulse, the so-called Lambropoulos curse can play a role in the dynamics of photoionization~\cite{lambropoulos}. 
Namely, it may happen that the hydrogen-like ion is fully ionized during the ramp-up part of the laser pulse, before the field acquires its maximum strength. 
In contrast, when the envelope varies rapidly, the field reaches values large enough to permit  stabilization against ionization in very short time intervals. In this case, 
the target may survive undisturbed even up to the maximum field strength in the pulse~\cite{fedorov}.

As described above, our approach offers many advantages as compared to the case when the infinite plane-wave laser field is considered. However, it is still crucial to show 
that the unitarity of the problem is not violated while applying the approximations described in this paper. In other words, that the total probability of ionization~\eqref{theory15}
is always less than one. To do so, we calculate below $P_{\mathrm{ion}}$ using the Monte Carlo method. 

First, we rewrite Eq.~\eqref{theory15} such that
\begin{equation}
P_{\mathrm{ion}}=\frac{1}{\alpha^2m_{\rm e}c^2}\int_0^{2\pi}\dd \varphi_{\bm{p}}\int_{-1}^{1}\dd \cos\theta_{\bm{p}}\int_{m_{\mathrm{e}}c^2}^{\infty}\dd E_{\bm{p}}\, {\cal P}(\bm{p}),
\label{theory21}
\end{equation}
where ${\cal P}({\bm p})$ is defined by Eq.~\eqref{theory18}. To numerically evaluate the energy integral, its upper limit is substituted by a finite value, 
denoted as $m_{\mathrm{e}}c^2+E_{\mathrm{max}}$. It is chosen such that the probability distribution is negligible for photoelectron final energies larger than $E_{\rm max}$.
In our calculations, we set $E_{\mathrm{max}}=30$~keV. Now, by introducing the following change of variables,
\begin{equation}
\varphi_{\bm{p}}=2\pi\xi_1,\, \cos\theta_{\bm{p}}=2\xi_2-1,\, E_{\bm{p}}=m_{\mathrm{e}}c^2+E_{\mathrm{max}}\xi_3,
\label{theory22}
\end{equation}
the total probability of ionization becomes
\begin{equation}
P_{\mathrm{ion}}=\frac{4\pi E_{\mathrm{max}}}{\alpha^2m_{\rm e}c^2}\int_0^1 \dd\xi_1\dd\xi_2\dd\xi_3\, {\cal P}(\bm{p}),
\label{theory23}
\end{equation}
where we integrate over a unit cube.
This expression is now suitable to apply the Monte Carlo method with the uniformly distributed variables $\xi_i$ ($i=1,2,3$) (see, e.g., \cite{KKextra1,KMK2013}).

\begin{figure}
\begin{center}
\includegraphics[width=6.5cm]{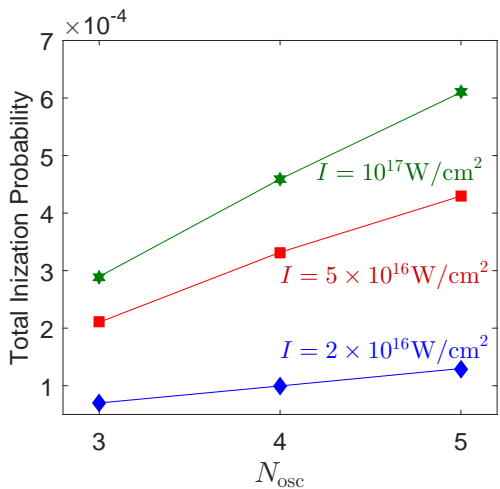}
\end{center}
\caption{Total ionization probabilities for pulses considered in this paper as functions of the number of cycles. The results of the Monte Carlo integration are plotted 
for three time-averaged intensities, as indicated in the figure. The points are connected with lines for graphical purposes. Note that, for the considered intensities, 
the total probability of ionization increases linearly with the number of field oscillations.
\label{mcplot}}
\end{figure}

In Fig.~\ref{mcplot}, we show the total probabilities of photoionization of He$^+$ ions by the laser pulse described in the previous Section. 
The data are for different time-averaged intensities such that $I=2\times 10^{16}$~W/cm$^2$ (diamonds), $I=5\times 10^{16}$~W/cm$^2$ (squares), and $I=1\times 10^{17}$~W/cm$^2$ (stars), 
and for different number of field cycles, $N_{\rm osc}$. The results were obtained from the Monte Carlo integration of Eq.~\eqref{theory23} 
by considering no less than $10^6$ sample points, with the estimated relative standard deviations smaller than 1\%. One can clearly see that
all calculated probabilities are less than one, which indicates the self-consistency of our treatment.
With this in mind, we can study now differential probability distributions of ionization and their properties.

\begin{figure}
\includegraphics[width=7cm]{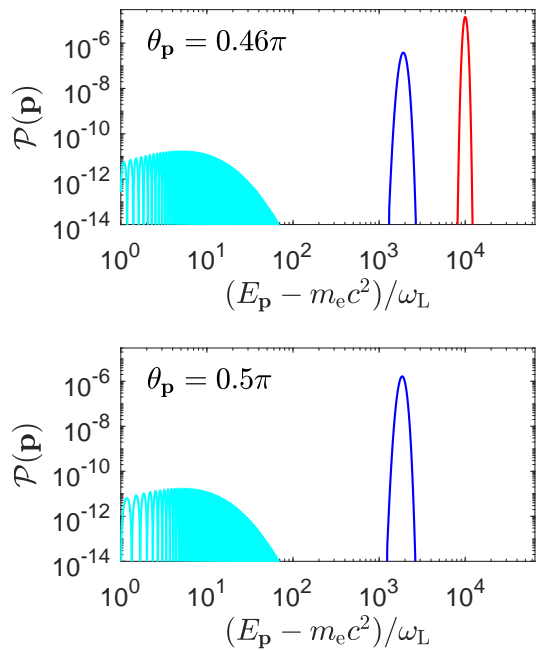}
\caption{Ionization probability distributions~\eqref{theory18} for the azimuthal angle $\varphi_{\bm{p}}=0.5\pi$ and two polar angles, $\theta_{\bm p}=0.46\pi$ (upper panel) and $\theta_{\bm p}=0.5\pi$ (lower panel). The laser pulse parameters are the same as in Fig.~\ref{figLaserFunctions160812}. 
 \label{figRIC2w2d161111}}
\end{figure}

\section{Energy probability distributions of ionization}
\label{energy_prob}

As it was mentioned before, the validity of the  RSFA is restricted to the case when the kinetic energy of photoelectrons is much larger than the ionization potential 
of the parent ion. Nevertheless, in this Section we are going to present the angular-resolved ionization probability distributions for photoelectron kinetic energies ranging 
from $0$ up to $10^5\omega_{\mathrm L}$. It is, therefore, important to keep in mind that the results corresponding to the low-energy part of the spectra provide only a qualitative insight into the process.

In Fig.~\ref{figRIC2w2d161111}, we present the respective energy distributions of ionized electrons when measured at fixed angles. Such distributions are obtained from Eq.~\eqref{theory18} 
by considering the interaction of a single He$^+$ ion with the laser pulse described in Fig.~\ref{figLaserFunctions160812}. While both panels relate to the same azimuthal detection angle, 
$\varphi_{\bm{p}}=0.5\pi$, they are for different polar angles: $\theta_{\bm p}=0.46\pi$ and $\theta_{\bm p}=0.5\pi$ for the upper and lower panel, respectively. In the upper panel, 
we observe three distinctive structures in the spectrum. The low-energy part (cyan curve), that spans the region from 0 to 60~eV, is the least pronounced and consists of fast oscillations. 
This is in contrast to the remaining structures. They appear as big lobes with maxima centered at either 2.96~keV in the case of the mid-energy pattern (blue curve) or 15.5~keV 
in the case of the high-energy one (red curve). The latter is by nearly two orders of magnitude more pronounced and, as we have checked for this particular
emission direction, it contributes the most to the energy-integrated 
probability distribution. In the lower panel of Fig.~\ref{figRIC2w2d161111}, which is for a slightly different polar angle, 
the high-energy structure is missing. Thus, the structure appears to be very sensitive to the polar angle $\theta_{\bm p}$. On the other hand, the low- and mid-energy patterns stay nearly the same. 
Such behavior can be related to radiation pressure, as we elaborate this next (for a discussion of radiation pressure, see, for instance, Ref.~\cite{KKpressure}).

\begin{figure}
\includegraphics[width=7cm]{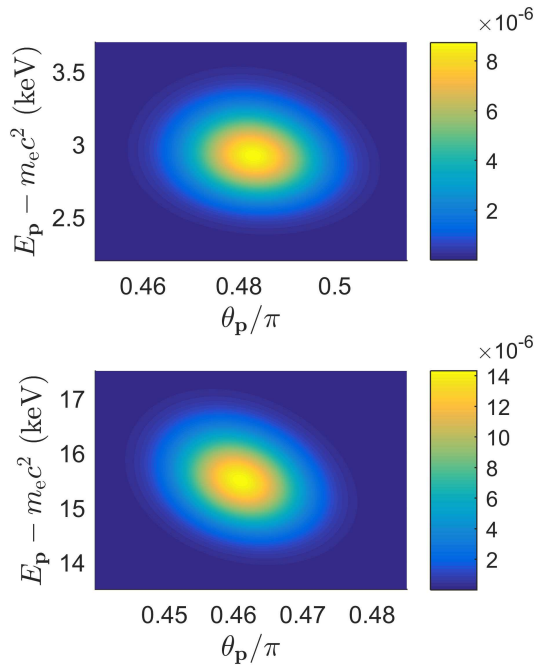}
\caption{Color mappings of ionization probability distributions ${\cal P}({\bm p})$ for the azimuthal angle $\varphi_{\bm{p}}=0.5\pi$. The intermediate- and high-energy structures are shown in the upper and lower panels, respectively. The laser field parameters are the same as in Fig.~\ref{figLaserFunctions160812}.
\label{figRelIon2ET161201}}
\end{figure}

Fig.~\ref{figRelIon2ET161201} shows the color mappings of the energy and polar angle distributions of photoelectrons from either the mid- (upper panel) or high-energy (lower panel)
regions obtained for $\varphi_{\bm p}=0.5\pi$. Both distributions have the cigar-like shape, even though they are displaced from each other. In the upper panel, the maximum of the distribution 
is located at the point $(\theta_{\bm{p}},E_{\bm{p}}-m_{\mathrm{e}}c^2)=(0.483\pi,2.92~\mathrm{keV})$ with the peak value $\mathcal{P}(\bm{p})=8.73\times 10^{-6}$. In the lower panel, it is 
located at $(0.461\pi,15.5~\mathrm{keV})$ with $\mathcal{P}(\bm{p})=14.3\times 10^{-6}$. Once again, we conclude that photoelectrons are mostly detected with high energies. 
Such energies are reached by absorption of many more laser photons than intermediate energies, leading to higher radiation pressure exerted on photoelectrons. This results in a displacement 
of the distribution towards smaller polar angles $\theta_{\bm p}$, i.e., in the direction of the driving pulse propagation~\cite{KKpressure}. Also, with absorption of a higher number of photons, 
the distribution becomes more spread in energy. While the width of the distribution in the upper panel is roughly 0.4~keV, in the lower panel it becomes 1~keV. On the other hand, the momentum 
transfer from many more photons makes the electron distribution more elongated in a certain direction. For this reason, the width of the distribution in $\theta_{\bm p}$ decreases from 
roughly $0.02\pi$ to $0.01\pi$ (upper and lower panels, respectively). This explains a strong sensitivity of the high-energy structure to the polar angle $\theta_{\bm p}$, observed in 
Fig.~\ref{figRIC2w2d161111}.

It has been shown in Ref.~\cite{no_interference} that the saddle-point analysis of the integrals in Eq.~\eqref{theory6} can give some insight into the mechanism of photoionization. 
Therefore, in the next Section, we introduce the saddle-point analysis to interpret our numerical results.

\subsection{Saddle-point analysis}
\label{saddle_point}

In order to perform the saddle-point analysis, the spin-resolved probability amplitude of ionization~\eqref{theory6} is written in terms of the light-cone variables. 
By doing so, as before, we are able to simplify Eq.~\eqref{theory6} to a single integral (see, also Ref.~\cite{no_interference}). Namely, we obtain
\begin{equation}
\mathcal{A}_{\lambda\lambda_{\mathrm{i}}}(\bm{p})=\frac{1}{k^0}\int_0^{2\pi}\dd\phi\,\ee^{\ii G(\phi)}\bigl[M_{\lambda\lambda_{\mathrm{i}}}(\phi)\bigr]_{\bm{q}=\bm{Q}},
\label{theory24}
\end{equation}
where $\phi=k^0x^-$, $M_{\lambda\lambda_{\mathrm{i}}}(\phi)$ is defined by Eq.~\eqref{theory7}, and it is calculated at the point ${\bm Q}={\bm p}+(q^0-p^0){\bm n}$. 
Additionally, the function $G(\phi)$ in Eq.~\eqref{theory24} is given by
\begin{align}
G(\phi)&\equiv G(g_0,g_1,g_2,h;\phi)=\int_0^{\phi}\dd\phi'  \bigl[g_0+g_1f_1(\phi')\nonumber\\
&+g_2f_2(\phi') +h\bigl(f_1^2(\phi')+f_2^2(\phi')\bigr)\bigr],
\label{theory25}
\end{align}
where
\begin{equation}
g_0=\frac{p^0-q^0}{k^0},  \;  h=\frac{(m_{\mathrm{e}}c\mu)^2}{2k\cdot p}, \; g_j=-m_{\mathrm{e}}c\mu\frac{\varepsilon_j\cdot p}{k\cdot p}.
\label{theory26}
\end{equation}

Now, given that the functions $G(\phi)$ and $\bigl[M_{\lambda\lambda_{\mathrm{i}}}(\phi)\bigr]_{\bm{q}=\bm{Q}}$ are sufficiently regular for real $\phi$ whereas $\ee^{\ii G(\phi)}$ 
is a fast oscillating function as compared to the remaining parts of the integrand in~\eqref{theory24}, the standard saddle-point approximation can be applied. The saddle points are obtained by solving the equation
\begin{equation}
\frac{{\rm d}G(\phi)}{{\rm d}\phi}=0.
\label{theory27}
\end{equation}
Among them, the ones that contribute to the integral in Eq.~\eqref{theory24}, from now on denoted as $\phi_s$, are those with $\textrm{Im}\,G(\phi_s)>0$ (or, equivalently, with ${\rm Im}\,\phi_s>0)$.
Keeping this in mind, the probability amplitude of ionization in the saddle-point approximation becomes
\begin{equation}
\mathcal{A}_{\lambda\lambda_{\mathrm{i}}}(\bm{p})=\frac{1}{k^0}\sum_s \ee^{\ii G(\phi_s)}\sqrt{\frac{2\pi\ii}{G^{\prime\prime}(\phi_s)}}\bigl[M_{\lambda\lambda_{\mathrm{i}}}(\phi_s)\bigr]_{\bm{q}=\bm{Q}}.
\label{theory28}
\end{equation}
This expression suggests that the interference-dominated structures in the energy spectra of photoelectrons should appear at regions for which two or more saddle points contribute the most 
to the above sum. In contrast, if just one of them dominates over a broad range of photoelectron kinetic energies [i.e., if $\textrm{Im}\,G(\phi_s)$ is considerably smaller compared 
to the corresponding values of the remaining saddle points], the interference-free structures are expected to be found. 

\begin{figure}
\begin{center}
\includegraphics[width=7.5cm]{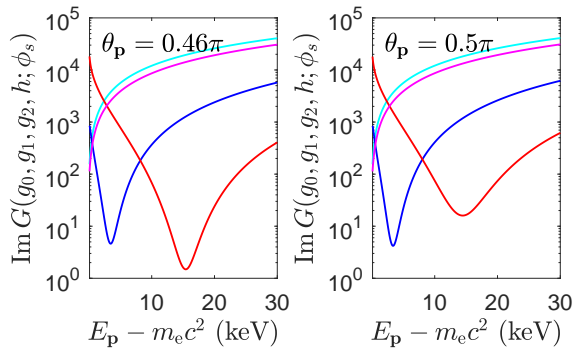}
\end{center}
\caption{Plots of $\mathrm{Im} G(g_0,g_1,g_2,h;\phi_s)$ for all contributing saddle points as functions of the photoelectron kinetic energy. The laser pulse parameters are the same as in 
Fig.~\ref{figLaserFunctions160812}. The azimuthal detection angle is chosen to be $\varphi_{\bm{p}}=0.5\pi$ whereas the polar angles are $\theta_{\bm{p}}=0.46\pi$ (left panel) and 
$\theta_{\bm{p}}=0.5\pi$ (right panel).
\label{figSaddleImG161111}}
\end{figure}

In Fig.~\ref{figSaddleImG161111}, we show the imaginary part of $G(\phi_s)$ as a function of the photoelectron kinetic energy for four saddle points which contribute to the sum in Eq.~\eqref{theory28}. 
While the laser field parameters are the same as in Fig.~\ref{figLaserFunctions160812} and $\varphi_{\bm{p}}=0.5\pi$ in both panels, two polar angles have been chosen: $\theta_{\bm{p}}=0.46\pi$ 
(left panel) and $\theta_{\bm{p}}=0.5\pi$ (right panel). Note that, for energies smaller than 60~eV, at least two saddle points contribute significantly to the probability amplitude of ionization~\eqref{theory28}
as the corresponding $\textrm{Im}\,G(\phi_s)$ takes the smallest values (cyan and magenta lines). As expected, this region coincides with the interference-dominated low-energy portion of the spectra presented 
in Fig.~\ref{figRIC2w2d161111}. Once again, we would like to stress that this is the energy domain beyond the validity of RSFA. For this reason, the analysis above can provide only a qualitative understanding 
of this particular interference pattern. Fortunately, for the considered laser pulse parameters and in contrast to the high-frequency case investigated in our previous 
studies~\cite{no_interference,proceedings_no_interference}, the low-energy electrons contribute marginally to the total ionization yield.

The interference-free structures are expected to appear at photoelectron kinetic energies at which an isolated saddle point contributes dominantly in Eq.~\eqref{theory28}. 
To illustrate this, we compare Figs.~\ref{figRIC2w2d161111} and~\ref{figSaddleImG161111}. For $\theta_{\bm p}=0.46\pi$, the mid- and high-energy interference-free structures peak 
in the energy region where $\textrm{Im}\,G(\phi_s)$ calculated at the dominant saddle points (blue and red curves in the left panel of Fig.~\ref{figSaddleImG161111}) reach their minimum values. 
The same conclusion can be drawn for $\theta_{\bm p}=0.5\pi$, with the difference that the high-energy lobe is absent in this case. To understand this we note that $\textrm{Im}\,G(\phi_s)$, 
shown as the red and blue curves in the left panel of Fig.~\ref{figSaddleImG161111}, have their minima which are by roughly three orders of magnitude smaller than the remaining values of 
$\textrm{Im}\,G(\phi_s)$ at the same electron energies. The situation in the right panel is changed as the minimal value of the red curve increases relative to the other values of $\textrm{Im}\,G(\phi_s)$. 
Thus, the high-energy structure disappears in the spectrum (see, Fig.~\ref{figRIC2w2d161111}). Note also that the values of $\textrm{Im}\,G(\phi_s)$ for the saddle point represented by the red curve seem to change rapidly 
with the angle $\theta_{\bm p}$. Thus, resulting in a sensitivity of the corresponding high-energy structure to the polar angle $\theta_{\bm p}$. This is in contrast to the blue curve, which
agrees with our earlier observation that the mid-energy lobe in Fig.~\ref{figRIC2w2d161111} is less sensitive to the change of $\theta_{\bm p}$.

The saddle-point analysis gives a reasonable estimate of the photoelectron time of emission, which is related to ${\rm Re}\phi_s$ (see, e.g., Ref.~\cite{no_interference}). Specifically,
the two saddle points causing the low-energy interference pattern in Fig.~\ref{figRIC2w2d161111} are such that, for the electron kinetic energies up to 60~eV, ${\rm Re}\phi_s$ 
changes as 0.28-0.38 (cyan line) and 5.78-5.87 (magenta line). This indicates that the corresponding photoelectrons are ionized at the beginning and at the end of the
driving pulse. The mid-energy interference-free structure, on the other hand, peaks at ${\rm Re}\phi_s=1.42$ (blue line). These electrons are emitted more towards the middle of the ramp-up
part of the pulse. Finally, the high-energy structure has its maximum at ${\rm Re}\phi_s=3.7$ (red line). Thus, the corresponding electrons are ionized
at roughly maximum value of the laser pulse envelope.

\begin{figure}
\begin{center}
\includegraphics[width=6cm]{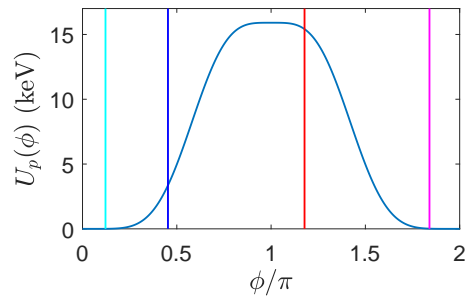}
\end{center}
\caption{Time-dependent ponderomotive energy [Eq.~\eqref{ponderomotive_energy}] of an electron in the laser field described in Fig~~\ref{figLaserFunctions160812}. 
The vertical lines are drawn for $\phi=0.38$ (cyan), $\phi=1.42$ (blue), $\phi=3.7$ (red), and $\phi=5.78$ (magenta). 
\label{figLaserFunctionsUp160812}}
\end{figure}

Finally, let us analyze the time-dependent ponderomotive energy of an electron undergoing the quiver motion in a laser field,
\begin{equation}
U_p(\phi)=\frac{e^2\bm{A}^2(\phi)}{2m_{\mathrm{e}}}\,.
\label{ponderomotive_energy}
\end{equation}
For the laser field defined in Fig.~\ref{figLaserFunctions160812}, $U_p$ is plotted in Fig.~\ref{figLaserFunctionsUp160812}. Hence, we see that photoelectrons which 
contribute to the mid- and high-energy structures appear in the continuum at times when the ponderomotive energy equals 3.36~keV and 15.41~keV (intersection of the blue 
and red vertical lines with the curve in Fig.~\ref{figLaserFunctionsUp160812}), respectively. Note that those values are close to the final photoelectron kinetic energies for which 
the interference-free distributions acquire their maximum values, i.e., 2.96~keV and 15.5~keV (see the discussion below Fig.~\ref{figRelIon2ET161201}). 
This suggests that, for the mid- and high-energy photoelectrons, the ponderomotive energy acquired at the moment of their birth in the continuum is basically transferred
into their longitudinal motion. Such an argument cannot be made for low-energy electrons. This discrepancy is 
actually expected, as the RSFA offers only a qualitative description of photoionization in this part of the spectrum.

Since the RSFA is particularly suitable to describe the highly energetic ionization, in the remaining part of this paper we focus on the high-energy interference-free 
structure that appears in the photoelectron spectrum of ionization. As we show next, it makes a potential for generating short-in-time electron wave packets.

\section{High-energy photoelectron wave packets}
\label{wave_packets}

To generate photoelectron wave packets, $\Psi_{\lambda\lambda_{\rm i}}(x)$, one has to superimpose the elementary electron waves, 
\begin{equation}
\Psi_{\lambda\lambda_{\mathrm{i}}}(x)=\int\frac{V\dd^3p}{(2\pi)^3}\sqrt{\frac{m_{\mathrm{e}}c^2}{VE_{\bm{p}}}}\,\ee^{-\ii p\cdot x}u_{\bm{p}\lambda}^{(+)}\mathcal{A}_{\lambda\lambda_{\mathrm{i}}}(\bm{p}),
\label{time1}
\end{equation}
with $\mathcal{A}_{\lambda\lambda_{\mathrm{i}}}(\bm{p})$ [Eq.~\eqref{theory12}] defining the profile of the wave packet. Note that the probability amplitude of
ionization $\mathcal{A}_{\lambda\lambda_{\mathrm{i}}}(\bm{p})$, that enters~\eqref{time1}, takes, in general, complex values. Therefore, the global phase of $\mathcal{A}_{\lambda\lambda_{\mathrm{i}}}(\bm{p})$ 
has to be a sufficiently regular function of ${\bm p}$ to guarantee that the plane waves in~\eqref{time1} interfere constructively.

\subsection{Global phase of the probability amplitude of ionization}

The spin-resolved probability amplitude of ionization $\mathcal{A}_{\lambda\lambda_{\mathrm{i}}}(\bm{p})$ can be represented as
\begin{equation}
\mathcal{A}_{\lambda\lambda_{\mathrm{i}}}(\bm{p})=\exp\bigl[\ii\Phi_{\lambda\lambda_{\mathrm{i}}}(\bm{p})\bigr]|\mathcal{A}_{\lambda\lambda_{\mathrm{i}}}(\bm{p})|,
\label{phase1}
\end{equation}
where the global phase equals $\Phi_{\lambda\lambda_{\mathrm{i}}}(\bm{p})=\mathrm{arg}[\mathcal{A}_{\lambda\lambda_{\mathrm{i}}}(\bm{p})]$. In our further analysis,
we will consider electron wave packets propagating in a given space direction determined by the fixed polar and azimuthal angles. For this reason, we will focus below on the energy dependence 
of the global phase, $\Phi_{\lambda\lambda_{\mathrm{i}}}(\bm{p})$, and the amplitude modulus squared, $|\mathcal{A}_{\lambda\lambda_{\mathrm{i}}}(\bm{p})|^2$, 
proportional to ${\cal P}_{\lambda\lambda_{\rm i}}({\bm p})$.
 
As mentioned above, in order to successfully synthesize an electron wave packet, the global phase of the ionization probability amplitude has to be regular. 
To quantify this statement, we expand the global phase in a Taylor series around the maximum of the high-energy distribution. Hence, we obtain a constant term, which 
is physically irrelevant, as well as the linear and quadratic terms, which introduce the time delay and chirp of the electron wave packet, respectively. 
In order to eliminate the chirp, the derivative of the global phase with respect to the photoelectron energy has to be nearly constant. Furthermore, this derivative should be 
spin-independent in order to guarantee that the time delay of the electron wave packet does not depend on the initial and final spin degrees of freedom.  
Note that, in our analysis, the initial spin state is assumed to be projected on the direction of laser pulse propagation, whereas the final spin state is projected on the direction 
of electron propagation, i.e., it is defined as the helicity.

\begin{figure}
\includegraphics[width=6.5cm]{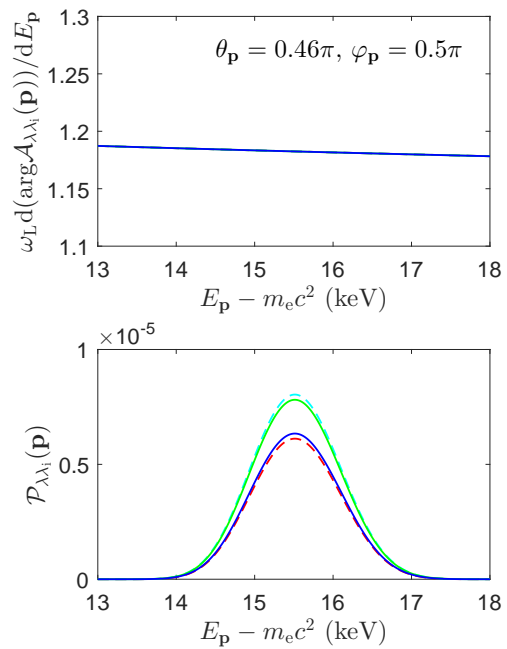}
\caption{Derivative of the global phases $\Phi_{\lambda\lambda_{\mathrm{i}}}({\bm{p}})$ with respect to the energy (upper panel) and the corresponding probability distributions, 
$\mathcal{P}_{\lambda\lambda_{\mathrm{i}}}(\bm{p})$ (lower panel), as functions of the photoelectron kinetic energy. 
The laser field parameters are the same as in Fig.~\ref{figLaserFunctions160812} and the electron emission angles are indicated in the upper panel. The four lines correspond 
to different spin degrees of freedom of the initial and final electron $(\lambda,\lambda_{\mathrm{i}})$: $(-,-)$ (dashed red line), $(+,-)$ (dashed cyan line), $(-,+)$ (solid green line), 
and $(+,+)$ (solid blue line). Note that, in the upper panel, all lines overlap.
\label{figPhaseDiff160812r1w3}}
\end{figure}

Fig.~\ref{figPhaseDiff160812r1w3} shows the derivative of the global phase $\Phi_{\lambda\lambda_{\mathrm{i}}}({\bm{p}})$ with respect to the electron energy (upper panel) and the spin-fixed probabilities 
of ionization, Eq.~\eqref{theory17} (lower panel), as functions of the photoelectron kinetic energy. The laser field parameters are the same as in Fig.~\ref{figLaserFunctions160812}, whereas 
the polar and azimuthal detection angles are $\theta_{\bm p}=0.46\pi$ and $\varphi_{\bm p}=0.5\pi$, respectively. The four curves correspond to different initial and final spin degrees of 
freedom of the electron $(\lambda,\lambda_{\mathrm{i}})$: $(-,-)$ (dashed red line), $(+,-)$ (dashed cyan line), $(-,+)$ (solid green line), and $(+,+)$ (solid blue line). 
In the upper panel, all curves overlap, meaning that the derivatives of all phases are spin-independent. We see that these derivatives are nearly
constant over a broad range of the final electron kinetic energies. This indicates that no chirp should be observed in the synthesized electron wave packets.
Note also that all spin-resolved probabilities shown in the lower panel are comparable, even though the spin-flipping processes are preferable.

\subsection{Space-time distributions}

Keeping in mind the above discussion, we go back to Eq.~\eqref{time1}. Now, if we want to analyze the electron wave packet propagating in a given space direction ${\bm n}_0$, 
defined by the polar and azimuthal angles $\theta_0$ and $\varphi_0$, respectively, we have to introduce in~\eqref{time1} constrains on possible values of
${\cal A}_{\lambda\lambda_{\rm i}}({\bm p})$. This can be done by multiplying ${\cal A}_{\lambda\lambda_{\rm i}}({\bm p})$ by the properly chosen filter function~\cite{no_interference},
\begin{align}
\mathcal{F}(\bm{p})=&\theta(E_{\mathrm{max}}-E_{\bm{p}}+m_{\mathrm{e}}c^2) \nonumber \\
 &\times\theta(E_{\bm{p}}-m_{\mathrm{e}}c^2-E_{\mathrm{min}})\delta^{(2)}(\Omega_{\bm{p}}-\Omega_{\bm{n}_0}).
\label{time1a}
\end{align}
In doing so, we create the electron wave packet that propagates in the direction $\bm{n}_0$ (with $\theta_0=\theta_{\bm{p}}$ and $\varphi_0=\varphi_{\bm{p}}$) 
and contains a range of free-electron energies $E_{\bm p}$, namely, $E_{\bm{p}}\in [E_{\mathrm{min}}+m_{\mathrm{e}}c^2,E_{\mathrm{max}}+m_{\mathrm{e}}c^2]$. This reduces the 
three-dimensional integral in Eq.~\eqref{time1} to the one-dimensional integral only. Thus, the space-time probability amplitude of ionization~\eqref{time1} becomes 
\begin{align}
\tilde{\mathcal{A}}_{\lambda\lambda_{\mathrm{i}}}(t,d)=&N_{\mathcal{A}}\int_{E_{\mathrm{min}}+m_{\mathrm{e}}c^2}^{E_{\mathrm{max}}+m_{\mathrm{e}}c^2}\dd E_{\bm{p}}
\ee^{-\ii (E_{\bm p} t-|{\bm p}|d) }u_{|{\bm p}|\bm{n}_{\bm{p}},\lambda}^{(+)} \nonumber \\
&\times \sqrt{E_{\bm{p}}}|\bm{p}|\mathcal{A}_{\lambda\lambda_{\mathrm{i}}}(|\bm{p}|\bm{n}_0),
\label{time2}
\end{align}
where the factor $N_{\mathcal{A}}$ contains all irrelevant constants and $d=\bm{n}_0\cdot\bm{x}$ is the electron distance from the parent ion. In the following, we extract from $\tilde{\mathcal{A}}_{\lambda\lambda_{\mathrm{i}}}(t,d)$ the common phase factor $\ee^{-\ii m_{\mathrm{e}}c^2 t}$ and define
\begin{equation}
\mathcal{A}_{\lambda\lambda_{\mathrm{i}}}(t,d)=\ee^{\ii m_{\mathrm{e}}c^2 t}\tilde{\mathcal{A}}_{\lambda\lambda_{\mathrm{i}}}(t,d).
\label{time3}
\end{equation}
This time-dependent factor artificially introduces rapid oscillations into the probability amplitude and is irrelevant when calculating probabilities.

Note that $\mathcal{A}_{\lambda\lambda_{\mathrm{i}}}(t,d)$ is a four-component object due to the presence of the free-electron bispinor $u_{\bm{p}\lambda}^{(+)}$. However, in the following, we will concentrate 
on its first component, $\mathcal{A}_{\lambda\lambda_{\mathrm{i}}}^{(1)}(t,d)$. This is because the space-time probability distributions, defined for each component $j=1,...,4$ as $\mathcal{P}_{\lambda\lambda_{\mathrm{i}}}^{(j)}(t,d)=|\mathcal{A}_{\lambda\lambda_{\mathrm{i}}}^{(j)}(t,d)|^2$ 
and with their maximum normalized to 1, are nearly identical, although the phases of $\mathcal{A}_{\lambda\lambda_{\mathrm{i}}}^{(j)}(t,d)$ are different. 

\begin{figure}
\includegraphics[width=7cm]{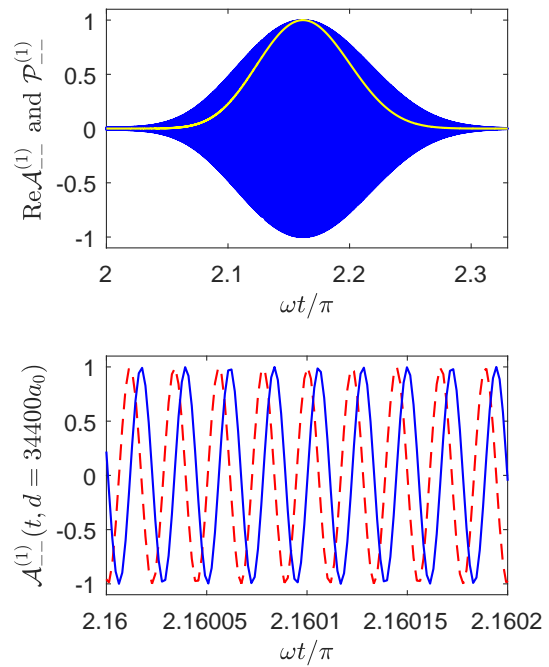}
\caption{First component of the space-time probability amplitude calculated at a distance $d=34400a_0$ from the parent ion for the case when the initial electron spin
is antiparallel to the propagation direction of the laser pulse and its final helicity has a negative projection, $\mathcal{A}^{(1)}_{--}(t,d=34400a_0)$. While in the 
upper panel only the real part of the probability amplitude is plotted (blue-filled area), in the lower panel the real (solid blue line) and the imaginary (dashed red line) parts are shown.
Note very fast oscillations of the probability amplitude, which on the scale presented in the upper panel occur as a blue area. Since the real and imaginary parts of 
$\mathcal{A}^{(1)}_{--}(t,d=34400a_0)$ are shifted in phase by roughly $\pi/2$, the corresponding probability $\mathcal{P}^{(1)}_{--}(t,d=34400a_0)$, plotted in the upper panel
as a solid yellow line, does not oscillate in time.
\label{figSpaceTimeAmp161202r1w3}}
\end{figure}

In Fig.~\ref{figSpaceTimeAmp161202r1w3}, we analyze the first component of the ionization probability amplitude~\eqref{time3} for the case when the initial and final electron spins
are antiparallel to the laser pulse and the electron propagation directions, respectively; in our notation, $\mathcal{A}^{(1)}_{--}(t,d)$. The results have been obtained
for a distance $d=34400a_0$ from the parent ion, where $a_0$ is the Bohr radius. This particular distance has been chosen such that the electron wave packet has just left the laser 
pulse, i.e., $\phi=k\cdot x\approx\omega t> 2\pi$. Moreover, we have taken $E_{\rm min}=7$~keV and $E_{\rm max}=18$~keV. Note that $\mathcal{A}^{(1)}_{--}(t,d=34400a_0)$ is a rapidly oscillating function of $t$, which is illustrated in the upper panel
by plotting its real part. To show these fast oscillations more clearly, in the lower panel, we plot both real (solid blue) and imaginary (dashed red) parts of $\mathcal{A}^{(1)}_{--}(t,d=34400a_0)$ 
in a very short time interval. They present either the sine or cosine type of behavior, that lead to a smooth oscillation-free behavior of the corresponding probability of ionization, 
$\mathcal{P}^{(1)}_{--}(t,d=34400a_0)$. The latter is presented in the upper panel as a solid yellow line.

Note that, even after extracting the phase related to the rest energy of the electron from the probability amplitude [cf. Eq.~\eqref{time3}], the latter is still a rapidly oscillating function
of time. This follows from the fact that the electron wave packet is built up from the high-energy interference-free portion of the spectrum, for which the condition 
$E_{\bm{p}}-m_{\mathrm{e}}c^2 \gg \omega$ is met. Furthermore, we have determined that the resulting electron pulse is very long, as it comprises at least few thousands of oscillations. 
It is known from laser physics that, for long pulses, the so-called carrier-envelope phase (CEP) is irrelevant. Hence, the same can be expected for the electron diffraction experiments.
It is anticipated that the change of CEP for the remaining components of the probability amplitude will not play a significant role either. Finally, we have estimated the width
of the electron pulse. As it follows from the upper panel of Fig.~\ref{figSpaceTimeAmp161202r1w3}, just after leaving the laser focus and before spreading in time, the electron pulse width
is roughly $\Delta t\approx\pi/(10\omega)\approx 400$~as.

\subsection{Analysis for larger intensities}

Consider a more intense laser field. It is expected that, for larger intensities of the driving laser field, the resulting high-energy portion of the photoelectron 
spectrum is shifted towards higher energies. This follows from the fact that the time-dependent ponderomotive energy~\eqref{ponderomotive_energy} increases linearly with intensity 
when the remaining laser field parameters are kept unchanged. Thus, for larger intensities, there is more energy transferred into the longitudinal motion of photoelectrons
(see, our analysis in Sec.~\ref{saddle_point}). It is not obvious, however, how the maximum and the width of this distribution change with increasing the
laser field intensity. For this reason, we consider now the case when the photoelectrons are released by the laser field of time-averaged intensity $I=10^{18}$~W/cm$^2$. 
In order to check the consistency of the RSFA for this laser intensity we have estimated, using the Monte Carlo method, that the total ionization probability  
equals $3.5\times 10^{-4}$ with the standard deviation smaller that $3.5\%$. This calculation was performed assuming that
$E_{\mathrm{max}}=200\,\mathrm{keV}$ in Eq.~\eqref{theory23}.

\begin{figure}
\includegraphics[width=7cm]{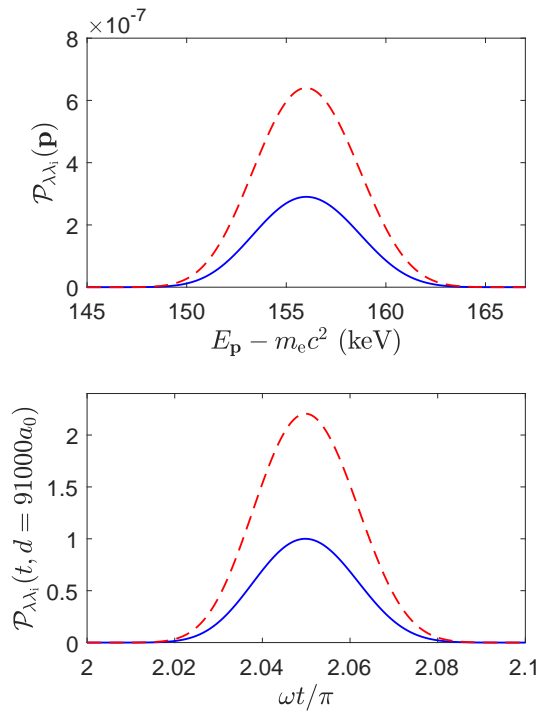}
\caption{ The spin-resolved probability distribution of ionization $\mathcal{P}_{\lambda\lambda_{\mathrm{i}}}(\bm{p})=|\mathcal{A}_{\lambda\lambda_{\mathrm{i}}}(\bm{p})|^2$ (upper panel) 
and the corresponding space-time probability of the synthesized electron wave packet $\mathcal{P}_{\lambda\lambda_{\rm i}}(t,d)$ (lower panel) calculated for the polar and azimuthal angles 
$\theta_{\bm{p}}=0.38\pi$ and $\varphi_{\bm{p}}=0.5\pi$, respectively. While the time-averaged intensity is $I=10^{18}\,\mathrm{W/cm}^2$, the remaining laser field parameters are the same 
as in Fig.~\ref{figLaserFunctions160812}. In both panels, we present the results for two possible spin configurations $(\lambda,\lambda_{\mathrm{i}})$: $(-,-)$ (solid blue lines) and 
$(-,+)$ (dashed red lines). Note that the space-time distributions are scaled to the maximum of $\mathcal{P}_{--}(t,d)$ and were calculated at a distance $d=91000a_0$ from the parent ion.
\label{figRelIon161203w6}}
\end{figure}

In the upper panel of Fig.~\ref{figRelIon161203w6}, we demonstrate the spin-resolved probability distribution of ionization~\eqref{theory17} calculated for the spin configurations $(\lambda,\lambda_{\mathrm{i}})$: $(-,-)$ 
(solid blue line) and $(-,+)$ (dashed red line) as a function of the photoelectron kinetic energy. While the averaged intensity of the laser pulse is $I=10^{18}$W/cm$^2$, its remaining  
parameters are the same as in Fig.~\ref{figLaserFunctions160812}. Moreover, the distributions are for the polar and azimuthal detection angles $\theta_{\bm{p}}=0.38\pi$ and
$\varphi_{\bm{p}}=0.5\pi$, respectively. As argued in Sec.~\ref{energy_prob}, the high-energy structure observed in the ionization probability distribution is very sensitive to 
the angle $\theta_{\bm p}$. By comparing Figs.~\ref{figRelIon2ET161201} and~\ref{figRelIon161203w6}, we observe that in the current case the high energy photoelectrons are detected at a much smaller 
polar angle than in Fig.~\ref{figRelIon2ET161201}. As the driving field is now more intense, it exerts a stronger radiation pressure on the photoelectrons. As a result, 
the respective probability distribution is shifted towards the direction of propagation of the laser field, i.e., towards smaller $\theta_{\bm p}$.
Moreover, by comparing $\mathcal{P}_{-\, -}(\bm{p})$ and $\mathcal{P}_{-\, +}(\bm{p})$ for $I=10^{18}$~W/cm$^2$ (upper panel of Fig.~\ref{figRelIon161203w6}) with the corresponding 
results for $I=10^{17}$~W/cm$^2$ (lower panel of Fig.~\ref{figPhaseDiff160812r1w3}), we see that the maximum of the probability distributions is now shifted by one order of magnitude towards 
larger photoelectron kinetic energies. The shifting is proportional to the increase of the laser pulse intensity (or the ponderomotive energy). Finally, we observe that the energy bandwidth is wider for the distributions presented in Fig.~\ref{figRelIon161203w6}
than for those in Fig.~\ref{figPhaseDiff160812r1w3}. This indicates that the electron pulses build up from these distributions should be shorter for the former.
To illustrate this, in the lower panel of Fig.~\ref{figRelIon161203w6}, we present the corresponding space-time probability distributions of the photoelectron wave packet 
$\mathcal{P}_{-\, -}(t,d)$ (solid blue line) and $\mathcal{P}_{-\, +}(t,d)$ (dashed red line), calculated at a distance $d=91000a_0$ from the parent ion. Both distributions, defined as
\begin{equation}
\mathcal{P}_{\lambda\lambda_{\mathrm{i}}}(t,d)=[\mathcal{A}_{\lambda\lambda_{\mathrm{i}}}(t,d)]^{\dagger}\mathcal{A}_{\lambda\lambda_{\mathrm{i}}}(t,d),
\label{totalspecetime}
\end{equation}
are scaled to the maximum of $\mathcal{P}_{--}(t,d=91000a_0)$. Now, the temporal width of the photoelectron pulse is roughly $\Delta t\approx 0.025\pi/\omega$, which is 
around four times shorter than for the case studied in Fig.~\ref{figPhaseDiff160812r1w3}. Note that the time duration of the electron pulse does not scale with 
the incident laser pulse intensity, as the position of the maximum of the probability distribution does. Actually, the ratio of the energy bandwidth to the energy position at which we observe the maximum 
decreases with increasing the laser intensity. In other words, the structure, if scaled to the most probable energy of emitted electrons, becomes more narrow.
As a consequence, the photoelectron pulse which leaves the laser focus lasts in the current case for roughly 100~as.

Note that the above discussion was based on the plane-wave-fronted pulse approximation for the driving laser field. Next, we will elaborate on the validity of this approximation
in relation to the laser pulse parameters used in this paper.

\subsection{Validity of the plane-wave-fronted pulse approximation}

Consider a laser pulse of power $P_{\mathrm{pulse}}=10$~PW, which is focused such that its time-averaged intensity distribution perpendicular to the laser pulse propagation is of the Gaussian form,
\begin{equation}
I(\bm{r}_{\bot})=I\exp\Bigl(-\frac{\bm{r}_{\bot}^2}{2\sigma_{\mathrm{L}}^2}\Bigr).
\label{val1}
\end{equation}
Here, $I$ is the laser beam peak intensity whereas $\sigma_{\mathrm{L}}$ determines the spatial size of the focus in the transverse direction. Thus, the total power of the pulse equals
\begin{equation}
P_{\mathrm{pulse}}=\int \dd^2r_{\bot}\, I\exp\Bigl(-\frac{\bm{r}_{\bot}^2}{2\sigma_{\mathrm{L}}^2}\Bigr)=2\pi\sigma_{\mathrm{L}}^2I.
\label{val2}
\end{equation}
If $I=10^{18}\,\mathrm{W/cm}^2$, then
\begin{equation}
\sigma_{\mathrm{L}}^2=\frac{P_{\mathrm{pulse}}}{2\pi I}=\frac{10^{-2}}{2\pi}\, \mathrm{cm}^2 \approx \frac{1}{(25)^2}\,\mathrm{cm}^2,
\label{val3}
\end{equation}
leading to $\sigma_{\mathrm{L}}\approx 4\times 10^{-4}\,\mathrm{m}$. We also estimate that for 10\,TW pulses of time-averaged intensity $I=10^{17}\,\mathrm{W/cm}^2$, 
the transverse size of the laser focus $\sigma_{\rm L}$ is by one order of magnitude smaller. In order to test the validity of the plane-wave-fronted pulse approximation, these values should 
be compared with the distance at which photoelectrons escape from the focus as measured in the perpendicular direction, $d^{\bot}_{\mathrm{escape}}$. Assuming that the position of the parent ion is not 
influenced by the laser field, due to the ion large mass, we estimate that $d^{\bot}_{\mathrm{escape}}$ is comparable to the separation between the electron wave packet and the
ion after leaving the laser focus, as the polar angle of emission is close to $\pi/2$. It is also expected that $d^{\bot}_{\mathrm{escape}}$ is the largest for the most energetic photoelectrons.

Our space-time analysis presented above shows that $d^{\bot}_{\mathrm{escape}}$ does not exceed $10^5\, a_0\approx 5\times 10^{-6}\,\mathrm{m}$ for the time-averaged intensity 
$I=10^{18}\,\mathrm{W/cm}^2$, and that it decreases for smaller intensities. Hence,
\begin{equation}
d^{\bot}_{\mathrm{escape}}\ll \sigma_{\mathrm{L}}.
\label{val4}
\end{equation}
In the current example, $d^{\bot}_{\mathrm{escape}}$ is smaller than the space-size of the laser focus by at least two orders of magnitude, provided that the total power of the pulse 
is around 10\,PW. Note that, for 10\,TW pulses and $I=10^{17}\,\mathrm{W/cm}^2$, the space-size of the focus is $\sigma_{\mathrm{L}}\approx 4\times 10^{-5}\,\mathrm{m}$ and 
$d^{\bot}_{\mathrm{escape}}\approx 4\times 10^{4}a_0\approx 2\times 10^{-6}\,\mathrm{m}$. This analysis shows that the plane-wave-fronted pulse approximation is suitable for the parameters 
chosen in the current paper. For lower pulse intensities, our estimations can be improved by increasing the total power of the pulse. On the other hand, for a fixed pulse power,
we can extend the validity of the plane-wave-front approximation by designing the laser focus which is not cylindrically symmetric but is elongated in a particular direction.
Finally, it is expected that this approximation can break for tighter focusing for which, however, the interference-free supercontinuum does not appear.

\section{Conclusions}
\label{conclusions}

Using the relativistic framework of strong-field approximation developed in Ref.~\cite{no_interference}, we have studied ionization of He$^+$ ions by intense laser pulses.
The latter have been treated in the plane-wave front approximation. We have shown that this approximation is suitable to describe ionization by not tightly focused pulses,
like the ones considered in this paper.

In addition to our earlier works~\cite{no_interference,proceedings_no_interference}, we have demonstrated a possibility of generating multiple supercontinua in the energy spectra of photoelectrons.
This is possible using relativistically intense laser pulses of circular polarization. Such intense pulses result in producing very energetic photoelectrons, whose 
final kinetic energy is practically determined by the ponderomotive energy that they acquire from the laser field at the moment of ionization. Specifically, for the parameters
used in this paper, we have shown that two supercontinua spanning the electron kinetic energies of keV and tens of keV are produced. Since the properties of a single supercontinuum
were already studied in Refs.~\cite{no_interference,proceedings_no_interference}, we have focused here on analyzing the conditions for observing the second supercontinuum.

As it has turned out, the high-energy supercontinuum is created in a very small polar-angular window. Moreover, it can be shifted towards even larger kinetic energies if one applies a more intense driving pulse. In this case, also its bandwidth
increases. As we have shown, this high-energy broad structure can lead to synthesis of attosecond electron pulses.

In general, photoelectron energy distributions crucially depend on the incident laser pulse parameters such as the frequency, number of cycles, polarization properties, etc. 
This is also clear when comparing the results presented in~\cite{no_interference,proceedings_no_interference} and in the current paper.
In Refs.~\cite{no_interference,proceedings_no_interference}, we have studied ionization by the high-frequency laser pulse and we have observed, for instance, that
the resulting high-energy lobes in the photoelectron energy spectrum are comparable in magnitude with the low-energy structures. On contrary, for a small-frequency laser pulse
considered in this paper, the corresponding low-energy structures are dramatically suppressed. When it comes to the position of supercontinua, 
they are determined by the temporal ponderomotive energy. As we show in this paper, their positions scale linearly
with the temporal ponderomotive energy. If we fix the averaged intensity of the laser pulse, the ponderomotive energy will be
larger for smaller pulse frequencies, pushing the supercontinua towards higher photoelectron energies. Therefore, one can expect that the most favorable 
conditions for the generation of high-energy supercontinuum and, hence, attosecond electron pulses are met for low-frequency laser fields.

In closing we note that, if ionization is driven by a finite train of laser pulses, one can expect to generate a finite sequence of electron pulses, similarly to the coherent combs 
investigated for the Thomson and Compton~\cite{KK1,KTK}, and for the Breit-Wheeler processes~\cite{KK2}. Let us also mention that interference effects related to ionization 
assisted by two (or more) laser pulses of different shapes and arbitrarily delayed with respect to each other (similar to the ones studied in~\cite{Martin2017} 
for the Breit-Wheeler process) can be also explored in the present context. Thus, ionization by relativistically intense laser pulses or by their trains 
can be used to engineer coherent electron pulses that are arbitrarily delayed and have different intensities. As the central 
energy of electron wave packets scales linearly with the time-averaged laser intensity, the generation of coherent electron beams in the MeV region can, in principle, be achieved 
in ELI and XCELS~\cite{ELI,XCELS}. In this context, the Compton effect studied theoretically in Ref.~\cite{DiPiazza2016} can be also investigated experimentally. Moreover, 
the sensitivity of the high-energy structures to the laser pulse parameters and to the ejection direction of photoelectrons can be used in the experimental diagnosis 
of extremely intense and short laser pulses, in addition to other methods exploring ionization spectra~\cite{Kalashnikov,KKion}. 
These topics are under investigations now.

\section*{Acknowledgements}

This work is supported by the National Science Centre (Poland) under Grant No. 2014/15/B/ST2/02203. We thank Antonino Di~Piazza for drawing our attention to the attoelectron diffraction physics
and to Martin Centurion for fruitful discussions about ultrashort electron pulse generation.

\end{document}